\documentclass[a4paper]{jpconf}
\usepackage{graphicx}
\begin{document}
\title{Edge states for the $n=0$ Laudau level in graphene}

\author{Mitsuhiro Arikawa$^1$, Yasuhiro Hatsugai$^1$, Hideo Aoki$^2$}

\address{$^1$ Institute of Physics,  University of Tsukuba, Tsukuba, 305-8571, Japan}

\address{$^2$Department of Physics, University of Tokyo, Hongo, Tokyo 113-0033, Japan}

\ead{arikawa@sakura.cc.tsukuba.ac.jp}

\begin{abstract}
In the anomalous quantum Hall effect (QHE), a hallmark of 
graphene, nature of 
the edge states in magnetic fields poses an important 
 question, since the edge and bulk should be intimately related in QHE.
Here we have theoretically studied the edge states, 
focusing on the $E=0$ edge mode, which is unusual in that 
the mode is embedded right within the $n=0$ bulk Landau level, 
while usual QHE edge modes reside across adjacent Landau levels.
Here we show that the $n=0$ Landau level, including the edge mode, 
has a wave function amplitude {\itshape accumulated} 
along zigzag edges whose width scales with the magnetic length, $l_B$.  
This contrasts with the usual QHE where 
the charge is depleted from the edge.  The implications are:
(i) The $E=0$ edge states in strong magnetic fields 
have a topological origin 
in the honeycomb lattice, so that they are 
outside the continuum (``massless Dirac") model.  
(ii) The edge-mode contribution decays only algebraically 
into the bulk, but this is ``topologically" compensated by 
the bulk contribution, resulting in the accumulation over $l_B$.  
(iii) The real space behavior obtained 
here should be observable in STM experiments.  
\end{abstract}

Ever since the anomalous quantum Hall effect (QHE) was experimentally 
observed,\cite{graphene1,graphene2} 
fascination with graphene is mounting.  
The interests have been focused on the ``massless Dirac" dispersions around Brillouin zone corners in 
graphene, where the Dirac cone is topologically protected 
due to the chiral symmetry\cite{topology1}.  
The peculiar dispersion is responsible for the appearance of the 
$n=0$ Landau level ($n$: Landau index) precisely around energy $E=0$ 
in magnetic fields\cite{fertig}.  For the ordinary integer QHE 
an important question 
is how the bulk and edge QHE conductions are related for finite samples.  
Many authors have 
addressed this question\cite{laughlin,halperin}, and 
the bulk QHE conductivity, a topological quantity, is shown to coincide 
with the edge QHE conductivity, itself another topological quantity.  
This is an example of the phenomena that, when a bulk system has a topological order\cite{topologicalorder1,topologicalorder2,topologicalorder3,
topologicalorder4} that reflects the geometrical phase of 
the system\cite{phase}, this should be reflected and become visible in the 
edge states in a bounded system.\cite{BEcorresponds1,BEcorresponds2}  
This ``bulk-edge correspondence" persists in graphene, as shown both 
from an analytic treatment of the topological numbers and numerical results for 
the honeycomb lattice\cite{topology2,dresden}.

In this paper, we reveal the features in 
the real-space profile 
of the edges states in graphene in magnetic fields $B$ 
in the one-body problem.  
A pecurier point on the graphene edge states is that the $E=0$ edge 
mode, despite being embedded right within the $n=0$ bulk Landau level 
in the energy spectrum, 
has a wave function whose charge is {\itshape accumulated} along zigzag edges.  This is drastically different from the ordinary QHE, where edge modes 
reside, in energy, between adjacent Landau levels, and 
their charge is depleted toward an edge.  
The physics here indicates a topological origin in a honeycomb lattice, which is in fact totally outside continuum models.  
In $B=0$, a zigzag edge in graphene 
has been known to have a flat dispersion at $E=0$,\cite{fujita} which 
is protected by the bipartite symmetry of the honeycomb lattice.  
Here we focus on the edge states in strong magnetic fields, 
which has a flat dispersion at $E=0$.

\begin{figure}
\begin{center}
\includegraphics[width=4.5in]{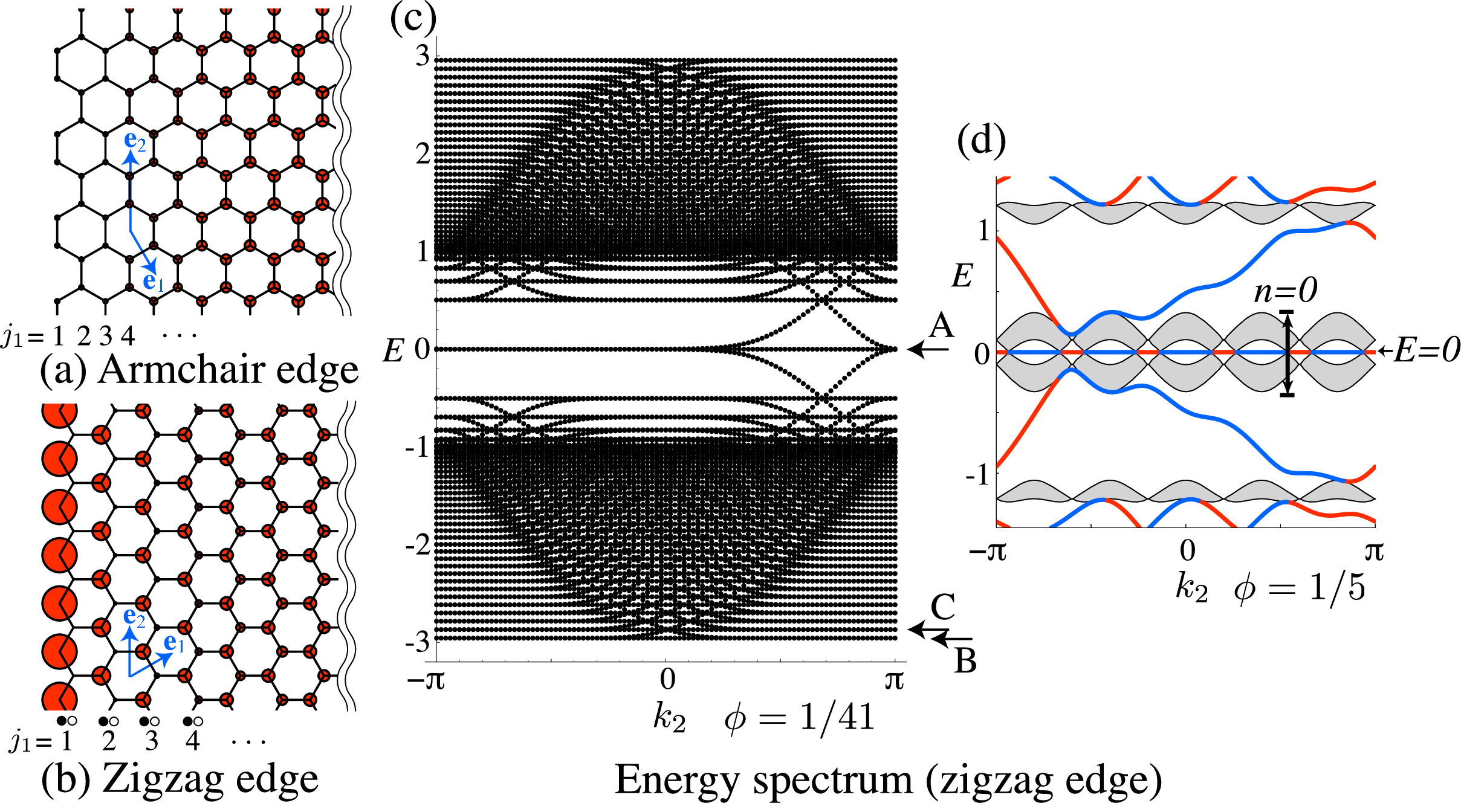}
\end{center}
\caption{
Honeycomb lattice with armchair(a) or zigzag(b) edges with 
${\mathbf e}_1, {\mathbf e}_2$ being respective unit translation vectors.
Local charge density ($\propto$ area of each circle) 
in a magnetic field $\phi=1/41$ 
for the energy window $-0.05<E<0.05$ around $n=0$ Landau 
level is displayed.    
(c) Energy spectrum for a zigzag edge against $k_2$ (momentum along the edge) 
in a magnetic field of $\phi=1/41$ for zigzag edges.  
(d) Blowup around the $n=0$ Landau level for a zigzag edge, 
in a magnetic field of $\phi=1/5$ here for clarity, 
with shaded regions represent the bulk energy spectra, while 
red curves the modes localized on the zigzag edge.
\label{fig1}  }
\end{figure}

We consider the tight-binding model on the honeycomb lattice 
with nearest-neighbor hopping $t$ (which is taken to the the unit of energy 
hereafter). The magnetic field is introduced as a Peierls phase.  
The flux in units of the magnetic flux quantum is
$
\phi \equiv  B S_6/(2\pi) = 1/q
$
in each hexagon with an area $S_6 = (3\sqrt{3}/2) a^2$.
Since honeycomb is a non-Bravais, bipartite lattice with two sublattice 
sites $\bullet$ and $\circ$ per unit 
cell, we can define two fermion operators 
$c_{\bullet}({\mathbf j})$ and $c_{\circ}({\mathbf j})$ 
with ${\mathbf j}= j_1{\mathbf e}_1+j_2{\mathbf e}_2$
defined in Figs.~\ref{fig1}(a) and(b) specifies the position of a unit cell.  
We assume that the spacing between the edges $L_1$($j_1=1,2,\cdots,L_1$) is taken to be large enough 
to avoid interference.
The length along the direction (${\mathbf e}_2$) parallel to the edge 
is also assumed to be long enough, 
for which we 
apply the periodic boundary condition. 
Performing a Fourier transform in that direction,
we obtain a $k_2$-dependent series of one-dimensional Hamiltonian, ${\mathcal H} = \sum_{k_2} {\mathcal H}_{1D}(k_2)$.
 The resultant eigenvalue problem reduces to 
${\mathcal H}_{1D}(k_2) | \psi(k_2,E) \rangle = E | \psi(k_2,E) \rangle $,
 with corresponding eigenstates 
 $|\psi(k_2,E) \rangle$. 

Having STM images in mind, we define the local charge density, 
\begin{eqnarray}
I(x(j_1)) & = &\frac{1}{2\pi}  \int ^{E_2}_{E_1} dE \int
d k_2 
|\psi_\alpha (E,j_1, k_2)|^2. 
\label{defchargedensity}
\end{eqnarray} 
Here $x$ is the distance from the edge 
(as related to $j_1$ via ${\mathbf e}_1$ which is not normal to the edge), 
and $E_1 < E < E_2$ is the energy window to be included in the 
charge density (which is normalized to unity when the window covers the 
whole spectrum).  

\begin{figure}
\begin{center}
\includegraphics[width=4.2in]{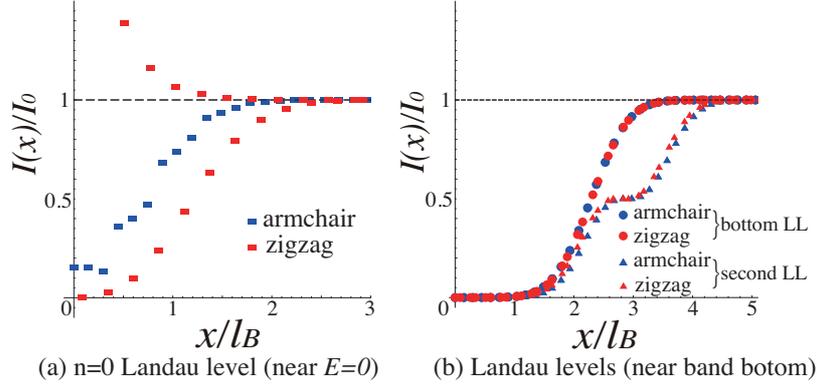}
\end{center}
\caption{(a) 
Scaled plot of the charge density $I(x)$ for magntic flux $\phi=1/41$ against $x/l_B$, the 
distance from the edge normalized by the magnetic length, 
for the $n=0$ Landau level (marked with A in Fig.~\ref{fig1}(c)).  
(b) The same plot for the outermost Landau levels (marked with B and C 
in Fig.~\ref{fig1}(c)), for which $I_0=\phi/2$.   \label{fig3}   }
\end{figure}

Figure~\ref{fig1}(c) shows the energy spectrum for a zigzag edge in a magnetic field $\phi=1/41$.  
For this magnetic field the $n=0$ Landau level around $E=0$, 
with a narrow energy width, almost looks like a 
line spectrum on this plot.   
We calculate the local charge density defined in 
eq.(\ref{defchargedensity}) for the armchair and zigzag edges.  
Figure~\ref{fig3}(a) depicts the charge density $I(x)$ normalized by 
the bulk value $I_0 (=\phi)$ 
against the distance from the edge $x$ measured by the magnetic length $l_B$ 
 for $\phi=1/41$, with the energy window $| E | <0.05$ 
 set to cover the $n=0$ Landau 
level (along with the embedded $E=0$ edge mode).  
The charge density for an armchair edge decreases 
monotonically toward the edge, 
where the depletion occurs 
on the magnetic length scale 
($l_B=3^{3/4}a/\sqrt{2\pi \phi})$), as in ordinary QHE systems.  
In sharp contrast, a zigzag edge has 
the charge density for the $\bullet$-sublattice that is 
{\it accumulated} toward the edge while the charge density for the 
$\circ$-sublattice is depleted.  
When we perform this scaled plot the charge density with various magnetic fields,
each of which coalesces on the common curves \cite{ourpaper}.

Outside the van Hove singurality, we recover the conventional 
Landau levels, so that we expect ordinary edge states.  
The result for the charge density for this energy region in 
Fig.\ref{fig3}(b) indeed shows that the charge density of outermost Landau 
levels (labelled 
B and C in Fig.\ref{fig1}) 
is depleted from the edge region over the magnetic length $l_B$ in usual 
fashion.  As expected, there is no difference in behavior between the 
A, B sublattices nor between armchair and zigzag edges.  

\begin{figure}
\begin{center}
\includegraphics[width=4.5in]{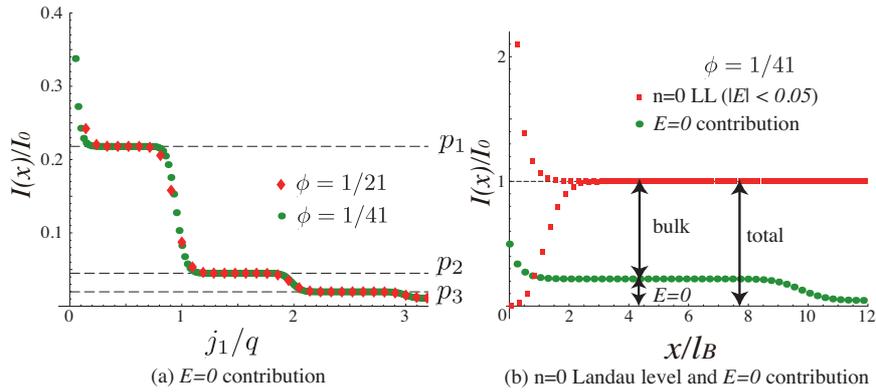}
\end{center}
\caption{Scaled plot of $I(x)$ at $E=0$ edge mode for $\phi(=1/q)=1/21$ and $1/41$ (a) and comparison with 
$n=0$ Landau Level contribution for $\phi=1/41$.  \label{fig3v}   }
\end{figure}

Going back to the $n=0$ Landau level for a zigzag edge, 
there is an important question: since the edge mode exists with an exactly 
$E=0$ flat dispersion, one might think that the charge accumulation 
along the edge entirely or primarily comes from this mode.  We have 
checked this.  
We can use the transfer-matrix method\cite{topology2} to examine the $I(x)$ 
contributed by the $E=0$ flat band, for which nonzero amplitudes occur, rigirously, only on  $\bullet$-sublattice, 
for a sample with an infinite length along the edge direction.  
Figure~\ref{fig3v}(a) presents $I(x(j_1))$ for $\phi=1/21$ and $1/41$, normalized by 
the bulk value of $n=0$ Landau level, $I_0 =\phi$.  
Interestingly enough, we can immediately notice a plateau structure.  
In this plot where the horizontal axis $j_1$ is normalized by $q 
(\propto 1/B)$, where 
all the data points for different magnetic fields fall upon a 
common curve, which means that the chage density contributed by the edge 
mode has a series of plateau structures with a step arising 
every time $j_1$ increases by $q$ for a magnetic field $\phi=1/q$.  
The height of the $n$-th plateau ($I(x)$ with $j_1=n$) 
can be analytically given in terms of  
$\bullet$-sublattice for the {\it zero} magnetic field, 
\begin{equation}
p_n \equiv \left.  I(x(n)) \right|_{\phi=0} 
=\frac{1}{\pi} \int_0^1  d t \frac{t^{2(n-1)}(1-t^2)}{\sqrt{1-t^2/4}}. 
\end{equation}
Asymptotically $p_n$ has an algebraic decay as 
$ p_n \simeq  {n^{-2}}/(\pi \sqrt{3})$.  
More interestingly, however, if we compare the total charge 
density with the contribution  from the $E=0$ flat band in 
Fig.~\ref{fig3v}(b) for the 
$n=0$ Landau level, the above plateau structure vanishes in the 
total density.  This implies that, although the edge-mode contribution 
has a slow, algebraic decay, the bulk contribution {\it compensates} 
this, and we end up with the charge accumulation over the magnetic 
length scale.  Since the Landau spectrum has a topological nature, 
we may call this curious phenomena a ``topological compensation of 
charge densities", which is the final key result here.  

To summarize, we have shown that the charge density of $n=0$ Landau level in 
graphene in strong magnetic fields should be 
totally unlike ordinary QHE systems.   
The charge accumulation along zigzag edges only occurs 
for the $E=0$ edge mode in the $n=0$ Landau level,  accompanied by 
a charge redistribution of the bulk states. 
The charge density around the edge can only be captured 
when  
bulk and edge contributions are considered simultaneously.  
The present result is expected to be measured by an STM imaging
for graphene edges\cite{stm}.   

\ack 
We wish to thank Hiroshi 
Fukuyama and Tomohiro Matsui 
for illuminating discussions.  
This work has been supported in part by Grants-in-Aid for Scientific Research, 
No.20340098, 20654034 from JSPS and 
No. 220029004 on Priority Areas from MEXT for MA, 
No.20340098, 20654034 from JSPS and 
No. 220029004, 20046002 on Priority Areas from MEXT for YH, 
No.20340098 from JSPS for HA.  


\end{document}